\documentclass[10pt,a4paper,twocolumn,english,aps,superscriptaddress,nofootinbib,nobibnotes,altaffillsymbol,showpacs,showkeys]{revtex4-1}
\usepackage[T1]{fontenc}
\usepackage[utf8]{inputenc}
\usepackage{lmodern}
\usepackage{color}
\usepackage{babel}
\usepackage{amsmath}
\usepackage{amssymb}
\usepackage{graphicx}
\usepackage{esint}
\usepackage[unicode=true,pdfusetitle,
 bookmarks=true,bookmarksnumbered=false,bookmarksopen=false,
 breaklinks=true,pdfborder={0 0 0},backref=false,colorlinks=true]
 {hyperref}
\hypersetup{
 linkcolor=BlueViolet,anchorcolor=BlueViolet,citecolor=BlueViolet,filecolor=BlueViolet,urlcolor=BlueViolet}

\makeatletter

\pdfpageheight\paperheight
\pdfpagewidth\paperwidth

 \@ifundefined{textcolor}{}
 {%
   \definecolor{BLACK}{gray}{0}
   \definecolor{WHITE}{gray}{1}
   \definecolor{RED}{rgb}{1,0,0}
   \definecolor{GREEN}{rgb}{0,1,0}
   \definecolor{BLUE}{rgb}{0,0,1}
   \definecolor{CYAN}{cmyk}{1,0,0,0}
   \definecolor{MAGENTA}{cmyk}{0,1,0,0}
   \definecolor{YELLOW}{cmyk}{0,0,1,0}
 }

\usepackage[usenames,dvipsnames]{xcolor}

\usepackage{titlesec}
\titlespacing*{\section}{0pt}{2em}{1em}
\titlespacing*{\subsection}{0pt}{1.5em}{1em}

\usepackage{bbm}

\@ifundefined{showcaptionsetup}{}{%
 \PassOptionsToPackage{caption=false}{subfig}}
\usepackage{subfig}
\makeatother

\begin{document}

\title{Simple QED- and QCD-like models at finite density}

\author{Jan M.~Pawlowski}

\affiliation{Institut für Theoretische Physik, Universität Heidelberg, Philosophenweg 16, 69120 Heidelberg, Germany}

\affiliation{ExtreMe Matter Institute EMMI, GSI, Planckstraße 1, D-64291 Darmstadt,
Germany}

\author{Ion-Olimpiu Stamatescu}

\affiliation{Institut für Theoretische Physik, Universität Heidelberg, Philosophenweg 16, 69120 Heidelberg, Germany}

\affiliation{FEST, Schmeilweg 5, 69118 Heidelberg, Germany}

\author{Christian Zielinski}

\thanks{Corresponding author.\\zielinski@pmail.ntu.edu.sg}

\affiliation{Institut für Theoretische Physik, Universität Heidelberg, Philosophenweg
16, 69120 Heidelberg, Germany}

\affiliation{Division of Mathematical Sciences, Nanyang Technological University,
Singapore 637371}

\date{30 July 2015}

\begin{abstract}
In this paper we discuss one-dimensional models reproducing some features
of quantum electrodynamics and quantum chromodynamics at nonzero
density and temperature. Since a severe sign problem makes a numerical
treatment of QED and QCD at high density difficult, such models help
to explore various effects peculiar to the full theory. Studying them
gives insights into the large density behavior of the Polyakov loop
by taking both bosonic and fermionic degrees of freedom into account,
although in one dimension only the implementation of a global
gauge symmetry is possible. For these models we evaluate the respective partition
functions and discuss several observables as well as the Silver Blaze
phenomenon.
\end{abstract}



\maketitle

\section{Introduction}

One of the open challenges of lattice gauge theory is the \textit{ab
initio} treatment of full quantum chromodynamics (QCD) at finite density
and low temperature. The fermion determinant is rendered complex and
rapidly oscillating after the introduction of a finite chemical potential
$\mu$. The same holds for a wide spectrum of theories at nonzero
density. The resulting near-cancellations make an evaluation of expectation
values extremely challenging. Several methods for meeting the sign
problem have been advanced, but are either limited in their applicability
or are about to be tested for QCD, see e.g.~Refs.~\cite{deForcrand:2010ys,Lombardo:2005gj,Aarts:2009yj,Aarts:2013uxa,Aarts:2013bla,Schmidt:2012uy,Mercado:2013yta,Alexandru:2005ix,Alexandru:2010yb,Langfeld:2012ah,Langfeld:2013xbf}.

In the past, studies of models of QCD have been proved insightful
\cite{Bilic:1987fn,Bilic:1988rw,Splittorff:2006fu,Splittorff:2007ck,Splittorff:2007zh,Ravagli:2007rw,Bloch:2013ara,Bloch:2013qva},
including in the special case of heavy quarks
\cite{DePietri:2007ak,Aarts:2008rr,Fromm:2012eb}. The interest
in the present and many other models discussed in the literature is
that they provide a testing ground for new simulation algorithms to
be applied to the full theory. Examples can be found in
Refs.~\cite{Aarts:2008rr,Seiler:2012wz,Bloch:2013ara,Bloch:2013qva}.

In this paper, we construct and study models that exhibit certain
characteristic properties of quantum electrodynamics (QED) and quantum
chromodynamics at nonzero $\mu$ and temperature $T$. These models
are formulated on a one-dimensional lattice using staggered fermions
\cite{Kogut:1974ag,Banks:1975gq,Banks:1976ia,Susskind:1976jm}. In
comparison to the full theories these models have several
simplifications. In particular, as it is not possible to define a
plaquette variable in one dimension, there is no Yang-Mills action and
the models can only respect a global
gauge symmetry. Nonetheless, the introduction of a bosonic field
allows us to go beyond models with only fermionic degrees of freedom.
Moreover, by the introduction of a suitable bosonic action $S_{\textrm{g}}$
we can mimic some features of Yang-Mills theory, which cannot
be directly translated to one dimension.

We note that our models generalize the one-link models presented in
Ref.~\cite{Aarts:2008rr}. With an explicit incorporation of bosonic degrees
of freedom with a corresponding action, a particular
form of the fermion matrix and an integration
over conjugacy classes we are able to construct a novel low-dimensional
QCD-like model,
which extends and cross-checks previous work.

As a result these models allow us to investigate some universal phenomena
also found in other models from a different perspective. In particular
our findings presented in Sec.~\ref{sec:QCD} shine new light on
the behavior of the Polyakov loop at large densities.

The resulting partition function can be fully integrated for a U(1)
gauge group. In the case of SU(3) we are left with an integral expression,
whose sign problem is manageable and which can be numerically evaluated.
We discuss some observables and investigate the Silver Blaze phenomenon
\cite{Cohen:2003kd} in these models. A theory exhibits Silver Blaze
behavior, if in the zero-temperature limit $T\to0$ observables become
independent of the chemical potential for $\mu\leq\mu_{\textrm{crit}}$,
where $\mu_{\textrm{crit}}$ is some critical onset value of the chemical
potential. In realistic models $\mu_{\textrm{crit}}=m_{q}^{\textrm{phys}}$
corresponds to the physical fermion mass, or, more generally, the
mass of the lowest excitation with nonvanishing quark number.

We organize the paper as follows: first we introduce a QED-like model
in Sec.~\ref{sec:QED}. We derive a closed expression for the partition
function and discuss the dependence of some observables on chemical
potential and temperature. In Sec.~\ref{sec:QCD} we deal with the
case of QCD and derive an integral expression, which can be numerically
evaluated. In Sec.~\ref{sec:Conclusions} we discuss and summarize
our findings.

\global\long\def\Tr{\operatorname{Tr}}

\global\long\def\diag{\operatorname{diag}}

\global\long\def\myRe{\operatorname{Re}}

\global\long\def\myIm{\operatorname{Im}}

\global\long\def\matrixOne{\mathbbm{1}}

\global\long\def\ii{\textrm{i}}

\section{A soluble, QED-like model at nonzero density \label{sec:QED}}

For the construction of a QED-like model we formulate an Abelian U(1)
lattice gauge theory on a finite one-dimensional lattice with staggered
fermions and couple them to a chemical potential, following a similar
ansatz as employed in previous works.

\subsection{Partition function}

\begin{figure*}[t]
\subfloat[Density, condensate and normalized susceptibility.]{\begin{centering}
\includegraphics[clip,angle=-90,width=0.99\columnwidth]{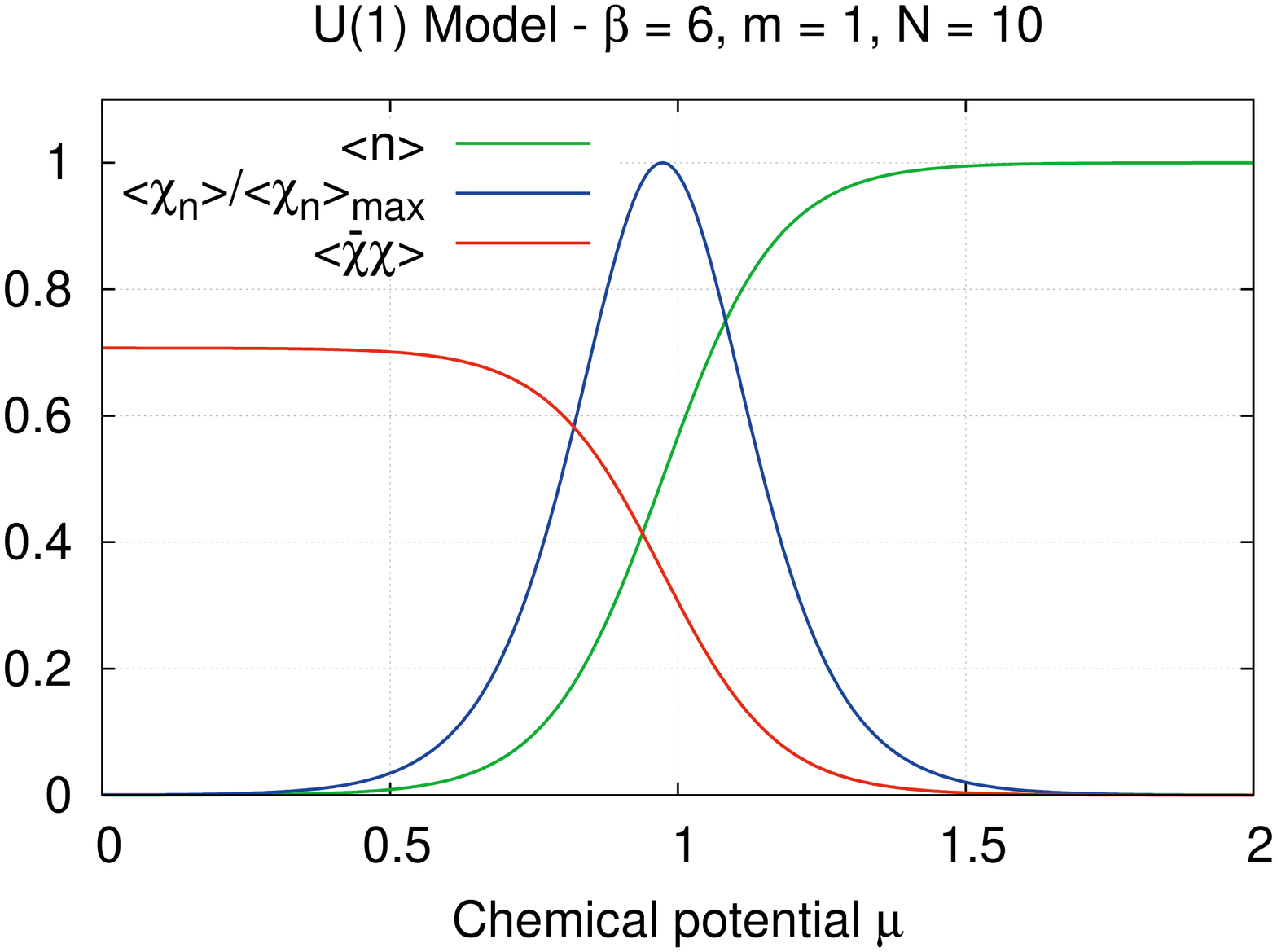} 
\par\end{centering}

}\hfill{}\subfloat[Polyakov loop and conjugate Polyakov loop.]{\begin{centering}
\includegraphics[clip,angle=-90,width=0.99\columnwidth]{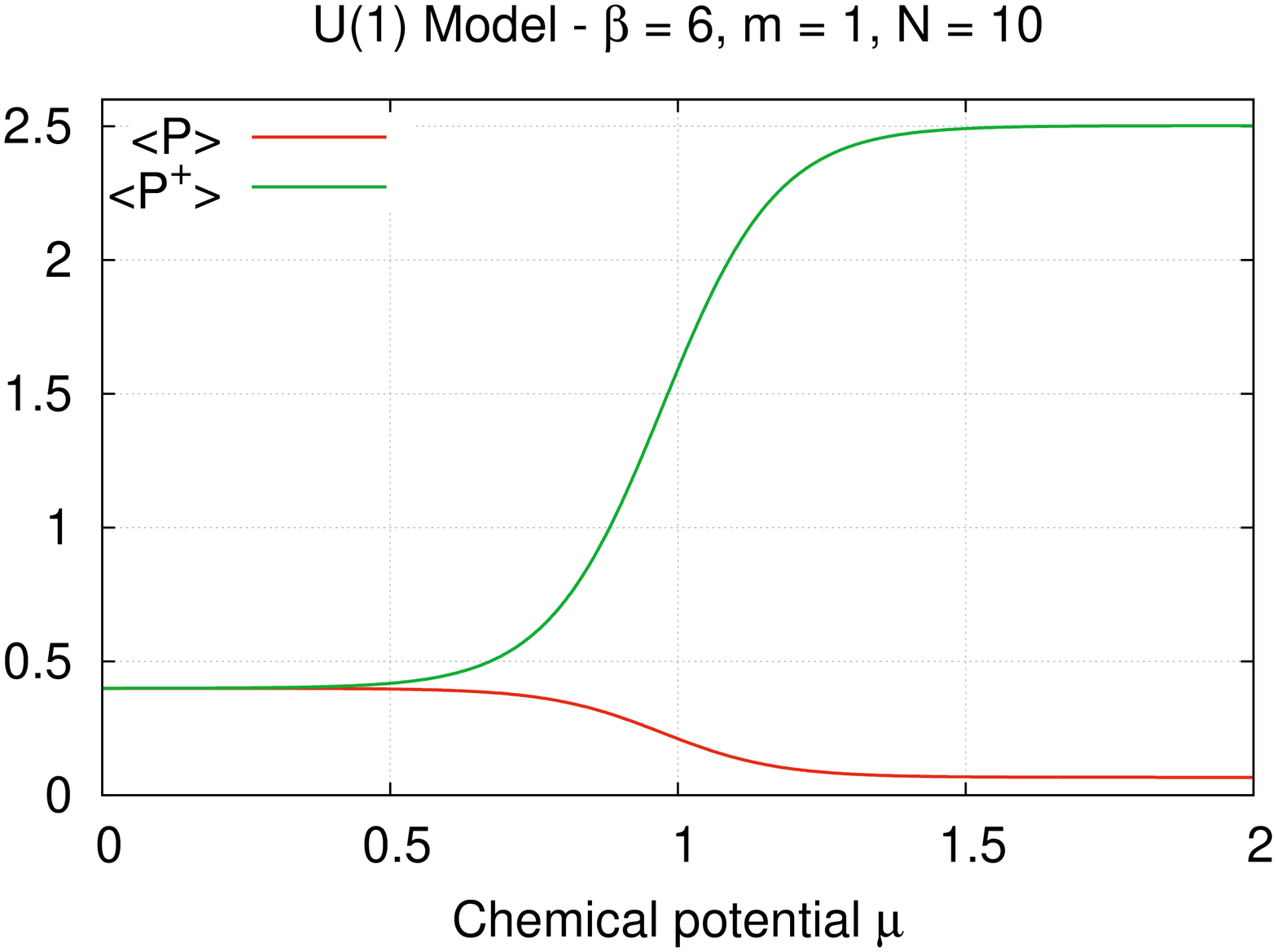} 
\par\end{centering}

}

\protect\caption{Observables in the U(1) model. \label{fig:Observables-U1}}
\end{figure*}

The action we employ mimics the usual compact lattice QED in one dimension.
The lattice is assumed to have a lattice spacing of $a$ and an extension
of $N$ sites, where $N$ is assumed to be even. We set $a=1$, i.e.,
we measure all dimensionful quantities in appropriate powers of $a$.
We consider a single staggered fermion field and couple it to a chemical
potential $\mu$. The temperature is identified with the inverse of
the lattice extension $T=N^{-1}$. The action $S=S_{\textrm{f}}+S_{\textrm{g}}$
consists of the fermionic part
\begin{equation}
S_{\textrm{f}}=\sum_{t,\tau=1}^{N}\overline{\chi}\left(t\right)K\left(t,\tau\right)\chi\left(\tau\right)\,,
\end{equation}
and the pure bosonic part
\begin{equation}
S_{\textrm{g}}=\beta\sum_{t=1}^{N}\left[1-\frac{1}{2}\left(U_{t}+U_{t}^{\dagger}\right)\right]\,.\label{eq:GaugeActionQED}
\end{equation}
Here $\overline{\chi}$ and $\chi$ denote the staggered fermion field,
$K\left(t,\tau\right)$ the fermion matrix, $\beta=1/e^{2}$ the inverse
coupling constant and $U_{t}\in\textrm{U(1)}$ the link variables.
The fermion matrix reads
\begin{equation}
K\left(t,\tau\right)=\frac{1}{2}\left(U_{t}e^{\mu}\delta_{t+1,\tau}-U_{\tau}^{\dagger}e^{-\mu}\delta_{t-1,\tau}\right)+m\delta_{t\tau}\,,
\end{equation}
with $m$ denoting the mass of the fermion. The introduction of the
chemical potential $\mu$ follows the prescription by Hasenfratz and
Karsch \cite{Hasenfratz:1983ba}. Furthermore we impose an antiperiodic
boundary condition for the fermionic field. After integrating out
the fermionic degrees of freedom, the partition function reads
\begin{equation}
Z=\intop\prod_{t=1}^{N}\textrm{d}U_{t}\,\det K\, e^{-S_{\textrm{G}}}\,.\label{eq:PartFunc}
\end{equation}
In this case the fermion determinant can be evaluated analytically
using identity (1) derived in Ref.~\cite{Molinari20082221}. We find
\begin{equation}
2^{N}\det K=e^{N\mu}\prod_{t}U_{t}+e^{-N\mu}\prod_{\tau}U_{\tau}^{\dagger}+2\rho_{+}\,,\label{eq:DetQED}
\end{equation}
where we have introduced
\begin{equation}
\rho_{\pm}=\lambda_{+}\pm\lambda_{-}\,,\qquad\lambda_{\pm}=\frac{1}{2}\left(m\pm\sqrt{1+m^{2}}\right)^{N}\,.
\end{equation}
Note that Eq.~\eqref{eq:DetQED}, like full QED, satisfies the identity
\begin{equation}
\det K\left(\mu\right)=\left[\det K\left(-\mu^{\star}\right)\right]^{\star}\,,\label{eq:DetSymmetry}
\end{equation}
which shows that in general the fermion determinant is complex for
$\mu>0$. We parametrize the link variables as $U_{t}=\exp\left(\ii\phi_{t}\right)$
in terms of algebra-valued fields $\phi_{t}\in\left[0,2\pi\right)$.
The corresponding U(1)-Haar measure reads
\begin{equation}
\intop\textrm{d}U_{t}=\int_{0}^{2\pi}\frac{\textrm{d}\phi_{t}}{2\pi}\,.
\end{equation}
With this parametrization, the action in Eq.~\eqref{eq:GaugeActionQED}
takes the form
\begin{equation}
S_{\textrm{g}}=\beta\sum_{t=1}^{N}\left(1-\cos\phi_{t}\right)\,.
\end{equation}
This allows us to integrate the partition function given in Eq.~\eqref{eq:PartFunc}
by using the expression we derived for the fermion determinant in
Eq.~\eqref{eq:DetQED}, to find
\begin{equation}
Z=\frac{e^{-\beta N}}{2^{N-1}}\left[\rho_{+}I_{0}^{N}\left(\beta\right)+\cosh\left(N\mu\right)I_{1}^{N}\left(\beta\right)\right]\,.\label{eq:QEDPartFunc}
\end{equation}
Here $I_{n}$ denotes the modified Bessel functions of the first kind.
As a cross-check we verified that $Z$ reduces to the previously
derived partition function in Ref.~\cite{Aarts:2008rr} for $N=1$ up to a
normalization constant, which depends on $m$ and $\beta$.

\subsection{Observables}

Given the final expression for the partition function in Eq.~\eqref{eq:QEDPartFunc},
we can easily calculate any observable of interest. The density follows
from $\left\langle n\right\rangle =N^{-1}\partial_{\mu}\log Z$, the
respective susceptibility is defined as $\left\langle \chi_{n}\right\rangle =\partial_{\mu}\left\langle n\right\rangle $
and the fermion condensate is given by $\left\langle \overline{\chi}\chi\right\rangle =N^{-1}\partial_{m}\log Z$.
For the density we then find,
\begin{equation}
\left\langle n\right\rangle =\frac{\sinh\left(N\mu\right)I_{1}^{N}\left(\beta\right)}{\rho_{+}I_{0}^{N}\left(\beta\right)+\cosh\left(N\mu\right)I_{1}^{N}\left(\beta\right)}\,.\label{eq:QEDDensity}
\end{equation}
Again this expression reduces to the known result in Ref.~\cite{Aarts:2008rr} for $N=1$. The fermion condensate follows as,
\begin{equation}
\left\langle \overline{\chi}\chi\right\rangle =\frac{\left(1+m^{2}\right)^{-1/2}\rho_{-}I_{0}^{N}\left(\beta\right)}{\rho_{+}I_{0}^{N}\left(\beta\right)+\cosh\left(N\mu\right)I_{1}^{N}\left(\beta\right)}\,.\label{eq:QEDFermCond}
\end{equation}
By directly evaluating the respective path integral expression, we
find for the Polyakov loop $\mathcal{P}=\prod_{t}U_{t}$ the expectation
value
\begin{equation}
\left\langle \mathcal{P}\right\rangle =\frac{e^{-\beta N}}{2^{N}Z}\left[2\rho_{+}I_{1}^{N}\left(\beta\right)+e^{N\mu}I_{2}^{N}\left(\beta\right)+e^{-N\mu}I_{0}^{N}\left(\beta\right)\right]\,.\label{eq:QEDPolyLoop}
\end{equation}
The conjugate Polyakov loop $\mathcal{P}^{\dagger}=\prod_{t}U_{t}^{\dagger}$
follows from a simple symmetry argument as $\left\langle \mathcal{P}^{\dagger}\right\rangle _{\mu}=\left\langle \mathcal{P}\right\rangle _{-\mu}$,
cf.~Ref.~\cite{Bloch:2013ara}.

In Fig.~\ref{fig:Observables-U1} we show the density given by Eq.~\eqref{eq:QEDDensity},
the fermion condensate by Eq.~\eqref{eq:QEDFermCond} and the Polyakov
loop by Eq.~\eqref{eq:QEDPolyLoop} as functions of the chemical
potential $\mu$. We see that already for $T=1/N=1/10$ the observables
only show a weak dependence on the chemical potential below some critical
value $\mu_{\textrm{crit}}$. This shows how the Silver Blaze behavior
\cite{Cohen:2003kd} becomes apparent in this model, which holds strictly
in the limit $T\to0$.

Close to $\mu\approx\mu_{\textrm{crit}}$ we also observe a fast increase
or decrease of the observables before reaching the saturation regime.
Note that the limits $\mu\rightarrow\infty$ and $\beta\rightarrow0$
do not commute. Density and condensate behave as one would expect.
The expectation values of $\mathcal{P}$ and $\mathcal{P}^{\dagger}$
approach 
\begin{equation}
\left\langle \mathcal{P}\right\rangle \to\left[\frac{I_{2}\left(\beta\right)}{I_{1}\left(\beta\right)}\right]^{N}\,,\qquad\left\langle \mathcal{P}^{\dagger}\right\rangle \to\left[\frac{I_{0}\left(\beta\right)}{I_{1}\left(\beta\right)}\right]^{N}\,,
\end{equation}
for $\mu\to\infty$. The Polyakov loop quickly drops to a typically
small value with increasing $\mu$ while the conjugate Polyakov loop
grows to a saturation value which diverges when $\beta\rightarrow0$.

\section{A QCD-like model at nonzero density \label{sec:QCD}}

Now we extend the previous model to the non-Abelian gauge group SU(3).
By restricting the integration over the full gauge group to the respective
conjugacy classes of SU(3), we will be able to reduce the partition
function to an integral expression with a manageable sign problem.

\subsection{Partition function}

\begin{figure*}[t]
\subfloat[Density and condensate.]{\begin{centering}
\includegraphics[clip,angle=-90,width=0.99\columnwidth]{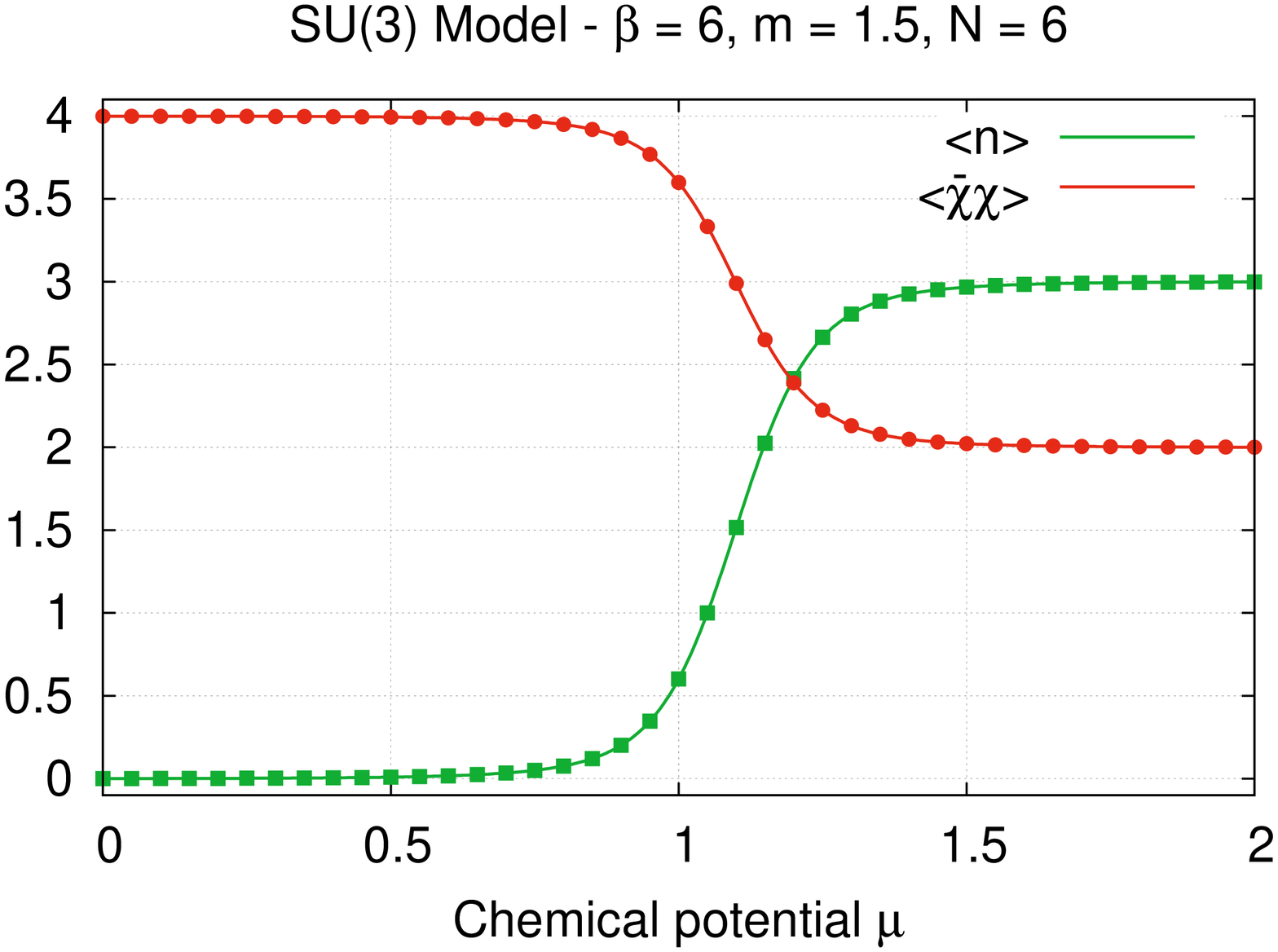} 
\par\end{centering}

}\hfill{}\subfloat[Polyakov loop and conjugate Polyakov loop.]{\begin{centering}
\includegraphics[clip,angle=-90,width=0.99\columnwidth]{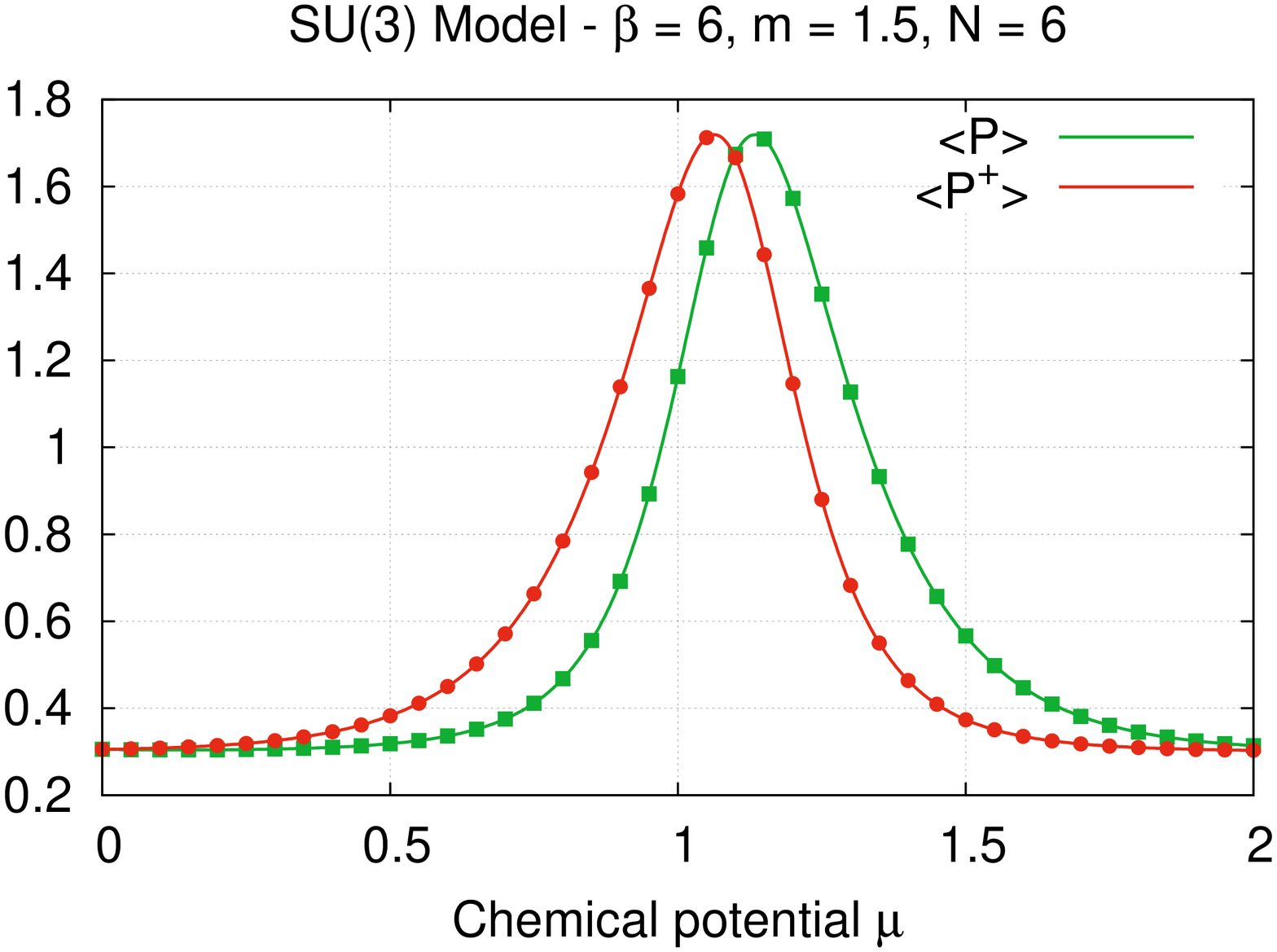} 
\par\end{centering}

}

\protect\caption{Observables in the SU(3) model. \label{fig:Observables-SU3}}
\end{figure*}

Our starting point is again the path integral expression for the partition
function in Eq.~\eqref{eq:PartFunc}, where now the pure bosonic
part of the action reads
\begin{equation}
S_{\textrm{g}}=\beta\sum_{t=1}^{N}\left[1-\frac{1}{6}\Tr_{\mathsf{c}}\left(U_{t}+U_{t}^{\dagger}\right)\right]\,.
\end{equation}
Here $\beta=6/g^{2}$ denotes the inverse coupling, $U_{t}\in\textrm{SU}\left(3\right)$
the link variables and $\Tr_{\mathsf{c}}$ a trace in color space.
Furthermore we replace the fermion matrix by
\begin{equation}
K\left(t,\tau\right)=\frac{1}{2}\left(\sigma_{+}U_{t}e^{\mu}\delta_{t+1,\tau}-\sigma_{-}U_{\tau}^{\dagger}e^{-\mu}\delta_{t-1,\tau}\right)+m\delta_{t\tau}\,,\label{eq:FermionMatrixQCD}
\end{equation}
with $\sigma_{\pm}=\frac{1}{2}\left(\matrixOne\pm\sigma_{3}\right)$
and the third Pauli matrix $\sigma_{3}=\diag\left(1,-1\right)$. In
the loop expansion this suppresses back steps, thus simulating a special
feature of Wilson fermions. This choice results in a factorization
of the fermion determinant of the form
\begin{equation}
\det K=\det\nolimits _{\mathsf{t,c}}\mathcal{K}_{f}\cdot\det\nolimits _{\mathsf{t,c}}\mathcal{K}_{b}\,,\label{eq:DetFactorQCD}
\end{equation}
where we introduced
\begin{align}
\det\nolimits _{\mathsf{t,c}}\mathcal{K}_{f} & =\det\nolimits _{\mathsf{t,c}}\left(m\delta_{t\tau}+\frac{1}{2}U_{t}^{\vphantom{\dagger}}e^{\mu}\delta_{t+1,\tau}\right)\,,\nonumber \\
\det\nolimits _{\mathsf{t,c}}\mathcal{K}_{b} & =\det\nolimits _{\mathsf{t,c}}\left(m\delta_{t\tau}-\frac{1}{2}U_{\tau}^{\dagger}e^{-\mu}\delta_{t-1,\tau}\right)\,.
\end{align}
Here $U_{N+1}=-U_{1}$ and $\det\nolimits _{\mathsf{t,c}}$ refers
to a determinant in position and color space.

In the following we restrict ourselves to observables which only depend
on the conjugacy class of the link variables. We then replace the
integration over the full gauge group SU(3) with an integration over
these conjugacy classes. This idea and the factorization given in
Eq.~\eqref{eq:DetFactorQCD} were also previously exploited in a
one link model in Ref.~\cite{Aarts:2008rr}. We thus parametrize
the links by
\begin{equation}
U_{t}=\diag\left(e^{\ii\phi_{t}},e^{\ii\vartheta_{t}},e^{-\ii\left(\phi_{t}+\vartheta_{t}\right)}\right)\,,\label{eq:ParametrizationSU3}
\end{equation}
with $\phi_{t},\vartheta_{t}\in\left(-\pi,\pi\right]$. Ignoring a
normalization constant, the Haar measure is given by $\textrm{d}U_{t}\propto J\left(\phi_{t},\vartheta_{t}\right)\,\textrm{d}\phi_{t}\,\textrm{d}\vartheta_{t}$
with
\begin{multline}
J\left(\phi_{t},\vartheta_{t}\right)=\sin^{2}\left(\frac{\phi_{t}-\vartheta_{t}}{2}\right)\\
\times\sin^{2}\left(\frac{\phi_{t}+2\vartheta_{t}}{2}\right)\sin^{2}\left(\frac{2\phi_{t}+\vartheta_{t}}{2}\right)\,,\label{eq:RedHaarMeas}
\end{multline}
while the bosonic part of the action takes the form
\begin{equation}
S_{\textrm{g}}=\beta\sum_{t=1}^{N}\left[1-\frac{1}{3}\left(\cos\phi_{t}+\cos\vartheta_{t}+\cos\left(\phi_{t}+\vartheta_{t}\right)\right)\right]\,.\label{eq:GaugeActionQCD}
\end{equation}
The determinant in position space has a simple structure and can be
analytically evaluated, e.g.~directly or by resummation of the loop
expansion for Eq.~\eqref{eq:FermionMatrixQCD}. For the remaining
determinant in color space we use the identity
\begin{equation}
\det\nolimits _{\mathsf{c}}\left(\matrixOne+\alpha U_{t}\right)=1+\alpha\Tr_{\mathsf{c}}U_{t}+\alpha^{2}\Tr_{\mathsf{c}}U_{t}^{-1}+\alpha^{3}\,,
\end{equation}
valid for all $A\in\textrm{SL}_{3}\left(\mathbb{C}\right)$, see Ref.~\cite{Aarts:2008rr}.
We can express the result in terms of the (conjugate) Polyakov loop
\begin{align}
\det\nolimits _{\mathsf{t,c}}\mathcal{K}_{f} & =m^{3N}\det\nolimits _{\mathsf{c}}\left(\matrixOne+\xi_{f}\prod_{t}U_{t}\right)\nonumber \\
 & =m^{3N}\left(1+\xi_{f}\mathcal{P}+\xi_{f}^{2}\mathcal{P}^{\dagger}+\xi_{f}^{3}\right)\,,
\end{align}
with $\xi_{f}=\left[\kappa\exp\left(\mu\right)\right]^{N}$ and hopping
parameter $\kappa=1/\left(2m\right)$. The Polyakov loop $\mathcal{P}$
and conjugate Polyakov loop $\mathcal{P}^{\dagger}$ are defined by
\begin{equation}
\mathcal{P}=\Tr_{\mathsf{c}}\prod_{t=1}^{N}U_{t}\,,\qquad\mathcal{P}^{\dagger}=\Tr_{\mathsf{c}}\prod_{t=1}^{N}U_{t}^{\dagger}\,.
\end{equation}
Analogously, we find
\begin{equation}
\det\nolimits _{\mathsf{t,c}}\mathcal{K}_{b}=m^{3N}\left(1+\xi_{b}\mathcal{P}^{\dagger}+\xi_{b}^{2}\mathcal{P}+\xi_{b}^{3}\right)\,,
\end{equation}
with $\xi_{b}=\left[\kappa\exp\left(-\mu\right)\right]^{N}$. We observe
that, as in the QED-like model, the fermion determinant in Eq.~\eqref{eq:DetFactorQCD}
satisfies the relation in Eq.~\eqref{eq:DetSymmetry}.

Putting all pieces together, we find that the partition function of
the model reads
\begin{align}
Z=m^{6N}\intop_{-\pi}^{\pi}&\left[\prod_{t}\textrm{d}\phi_{t}\,\textrm{d}\vartheta_{t}\, J\left(\phi_{t},\vartheta_{t}\right)\right]\nonumber\\
&\times\left(1+\xi_{f}\mathcal{P}+\xi_{f}^{2}\mathcal{P}^{\dagger}+\xi_{f}^{3}\right)\nonumber\\
&\times\left(1+\xi_{b}\mathcal{P}^{\dagger}+\xi_{b}^{2}\mathcal{P}+\xi_{b}^{3}\right)e^{-S_{\textrm{g}}}\,,\label{eq:PartFuncQCD}
\end{align}
where an irrelevant numerical normalization constant has been dropped.
The measure term $J\left(\phi_{t},\vartheta_{t}\right)$ was given
in Eq.~\eqref{eq:RedHaarMeas}, $S_{\textrm{g}}$ was introduced
in Eq.~\eqref{eq:GaugeActionQCD}.

\subsection{Observables}

Considering Eq.~\eqref{eq:PartFuncQCD} as a partition function for a
model of QCD, we can derive integral expressions for the
density, the susceptibility and the fermion condensate by taking corresponding
derivatives of $\log Z$. For the Polyakov loop and the conjugate
Polyakov loop we insert a $\frac{1}{Z}\mathcal{P}$ or a $\frac{1}{Z}\mathcal{P}^{\dagger}$
term in Eq.~\eqref{eq:PartFuncQCD}.

The resulting integral expressions are numerically evaluated. Typical
examples of these observables can be found in Fig.~\ref{fig:Observables-SU3}.
The density and the condensate show similar qualitative behavior to
the corresponding observables in the U(1) model, where we now find
$\left\langle n\right\rangle \to3$ for $\mu\to\infty$.

The Polyakov loop and conjugate Polyakov loop show some nontrivial
behavior. Close to the critical onset $\mu_{\textrm{crit}}$, we find
peaks in $\left\langle \mathcal{P}\right\rangle $ and $\left\langle \mathcal{P}^{\dagger}\right\rangle $
with the peak in the conjugate Polyakov loop appearing at smaller
$\mu$. Similar behavior was previously observed in a simulation of
a gauge theory with exceptional group $G_{2}$ \cite{Maas:2012wr},
a strong coupling limit in HQCD \cite{Seiler:2012wz}, a three-dimensional
effective theory of nuclear matter \cite{Fromm:2012eb} and in recent
studies of one-dimensional QCD \cite{Bloch:2013ara,Bloch:2013qva}.
The drop of the Polyakov loop at high density is easily understood
as an effect of saturation, while the displacement of the peaks has
a dynamical basis, see, e.g.~Ref.~\cite{Seiler:2012wz}.

Despite making use of a different approach, in general we find good qualitative
agreement with the results reported in Refs.~\cite{Bloch:2013ara,Bloch:2013qva}
after dropping $S_{\textrm{g}}$, i.e.\,for $\beta=0$.

\section{Conclusions \label{sec:Conclusions}}

In this paper we have constructed one-dimensional lattice models resembling
QED and QCD to investigate the finite density and finite temperature
regime. Despite the drastic simplifications in these models, they
capture some essential physical properties expected from the full
theory and show an interesting behavior of the Polyakov loop. We found
that they---like their four-dimensional continuous counterparts---exhibit
the Silver Blaze property in the zero temperature limit $N\to\infty$.
The $\mu$-dependence of the SU(3) (conjugate) Polyakov loop $\mathcal{P}$
($\mathcal{P}^{\dagger}$) shows the peculiar $\mu$-dependence also
found in other approximations of QCD. The models presented here can
also serve as a starting point for the construction of more elaborated
models.

\section{Acknowledgments}

This work is supported by the Helmholtz Alliance HA216/EMMI and by
ERC-AdG-290623. J.M.P. thanks the Yukawa Institute for Theoretical Physics,
Kyoto University, where this work was completed during the YITP-T-13-05
on 'New Frontiers in QCD'. I.-O.S. thanks the Deutsche Forschungsgemeinschaft
by STA 283/16-1 and C.Z. thanks Nanyang Technological University for support.

\bibliography{simple_models}

\end{document}